\documentclass[aps,prl,reprint,superscriptaddress,floatfix,showkeys]{revtex4-1}

\usepackage{graphicx}
\usepackage{amsmath}

\newcommand{\sv}{\ensuremath{\langle\sigma_{\text{ann}}v\rangle}}
\newcommand{\vmax}{V_{\rm max}}

\begin{document}


\title{Constraining Dark Matter Models from a Combined Analysis of Milky Way Satellites with the Fermi Large Area Telescope}

\author{M.~Ackermann}
\affiliation{W. W. Hansen Experimental Physics Laboratory, Kavli Institute for Particle Astrophysics and Cosmology, Department of Physics and SLAC National Accelerator Laboratory, Stanford University, Stanford, CA 94305, USA}
\author{M.~Ajello}
\affiliation{W. W. Hansen Experimental Physics Laboratory, Kavli Institute for Particle Astrophysics and Cosmology, Department of Physics and SLAC National Accelerator Laboratory, Stanford University, Stanford, CA 94305, USA}
\author{A.~Albert}
\affiliation{Department of Physics, Center for Cosmology and Astro-Particle Physics, The Ohio State University, Columbus, OH 43210, USA}
\author{W.~B.~Atwood}
\affiliation{Santa Cruz Institute for Particle Physics, Department of Physics and Department of Astronomy and Astrophysics, University of California at Santa Cruz, Santa Cruz, CA 95064, USA}
\author{L.~Baldini}
\affiliation{Istituto Nazionale di Fisica Nucleare, Sezione di Pisa, I-56127 Pisa, Italy}
\author{J.~Ballet}
\affiliation{Laboratoire AIM, CEA-IRFU/CNRS/Universit\'e Paris Diderot, Service d'Astrophysique, CEA Saclay, 91191 Gif sur Yvette, France}
\author{G.~Barbiellini}
\affiliation{Istituto Nazionale di Fisica Nucleare, Sezione di Trieste, I-34127 Trieste, Italy}
\affiliation{Dipartimento di Fisica, Universit\`a di Trieste, I-34127 Trieste, Italy}
\author{D.~Bastieri}
\affiliation{Istituto Nazionale di Fisica Nucleare, Sezione di Padova, I-35131 Padova, Italy}
\affiliation{Dipartimento di Fisica ``G. Galilei", Universit\`a di Padova, I-35131 Padova, Italy}
\author{K.~Bechtol}
\affiliation{W. W. Hansen Experimental Physics Laboratory, Kavli Institute for Particle Astrophysics and Cosmology, Department of Physics and SLAC National Accelerator Laboratory, Stanford University, Stanford, CA 94305, USA}
\author{R.~Bellazzini}
\affiliation{Istituto Nazionale di Fisica Nucleare, Sezione di Pisa, I-56127 Pisa, Italy}
% ...
\collaboration{The $Fermi$-LAT Collaboration}
\noaffiliation

\author{M.~Kaplinghat}
\affiliation{Center for Cosmology, Physics and Astronomy Department, University of California, Irvine, CA 92697-2575, USA}
\author{G.~D.~Martinez}
\affiliation{Center for Cosmology, Physics and Astronomy Department, University of California, Irvine, CA 92697-2575, USA}



\begin{abstract}

Satellite galaxies of the Milky Way are among the most promising targets for dark matter searches in gamma rays. We present a search for dark matter consisting of weakly interacting massive particles, applying a joint likelihood analysis to 10 satellite galaxies with 24 months of data of the Fermi Large Area Telescope. No dark matter signal is detected. Including the uncertainty in the dark matter distribution, robust upper limits are placed on dark matter annihilation cross sections. The 95\% confidence level upper limits range from about $10^{-26} \, {\rm cm}^3 {\rm s}^{-1}$ at 5 GeV to about $5\times10^{-23} \,{\rm cm}^3 {\rm s}^{-1}$ at 1 TeV, depending on the dark matter annihilation final state. For the first time, using gamma rays, we are able to rule out models with the most generic cross section ($\sim 3 \times 10^{-26} \, {\rm cm}^3 {\rm s}^{-1}$ for a purely s-wave cross section), without assuming additional boost factors.
\end{abstract}


\maketitle

\section{Introduction}
It is well-established that baryons contribute only about 20\% of the mass density of matter in the Universe \cite{Komatsu:2010fb}. The nature of the remaining 80\% of matter, known as dark  matter (DM), remains a mystery. One leading candidate consists of weakly interacting massive particles (WIMP), predicted in several extensions of the standard model of particle physics. If the WIMP is a Majorana fermion, its pair annihilation will produce gamma rays with a flux given by $\phi(E,\psi)=\sv/(8\pi m_W^2)\times N_W(E)\times J(\psi)$, where $\sv$ is the velocity averaged pair annihilation cross section, $m_W$ is the WIMP mass, $N_W(E)$ is the gamma-ray energy distribution per annihilation, and  $J(\psi)=\int_{l.o.s.,\Delta\Omega}dl \,d\Omega\rho^{2}[l(\psi)]$ is the line-of-sight (l.o.s.) integral of the squared DM density, $\rho$, toward a direction of observation, $\psi$, integrated over a solid angle, $\Delta\Omega$ (see {\em e.g.}, \cite{Bergstrom:1997fj}; see also \cite{Bergstrom:2000pn} for a review). 

Regions of local DM density enhancements with large $J(\psi)$, or J factors, are potentially good targets for DM searches. Dwarf spheroidal satellite galaxies (dSphs) of the Milky Way are DM-dominated systems without active star formation or detected gas content \cite{Mateo:1998wg,Grcevich:2009gt}. Thus, though the expected number of signal counts is not as high as from the Galactic center for instance, dSphs exhibit a favorable signal to noise ratio, and upper limits on a gamma-ray signal from DM annihilation have been obtained by the $Fermi$ Large Area Telescope ($Fermi$-LAT) \cite{Abdo:2010ex,Scott:2009jn} as well as air Cherenkov telescopes \cite{Aharonian:2007km,Albert:2007xg,:2010pja,Aleksic:2011jx}.  

In this Letter, we present new $Fermi$-LAT results on dSphs, with an updated dataset and two significant improvements over our previous analyses: first, we combine all the dSph observations into a single joint likelihood function, which improves the statistical power of the analysis, and second, we take into account the uncertainties in estimates of the J-factors, thereby making our results more robust.

\section{Fermi-LAT observations} 
The $Fermi$-LAT, the main instrument on board the $Fermi$ observatory, is a pair-conversion telescope that detects gamma rays in the energy range from $20$ MeV to $>300$ GeV with unprecedented sensitivity. Further details on the instrument can be found in \cite{Atwood:2009ez}, and current official performance figures are available in \cite{Perfpage}. 

In this Letter, we use 24 months of $Fermi$-LAT data, recorded between 2008-08-04 and 2010-08-04, and the data reduction is performed with the $Fermi$-LAT data analysis package, ScienceTools \cite{STpage}. Only \textquotedblleft diffuse\textquotedblright class events with energy between 200 MeV and 100 GeV are used. To avoid contamination from Earth limb gamma rays, events with zenith angles larger than $100^\circ$ are rejected and time intervals when the observed sky position is occulted by the Earth are discarded from the lifetime calculation. We extract from this dataset regions of interest (ROIs) of radius $10^\circ$ around the position of each dSph specified in Table~\ref{dwarfs}. 

\begin{table}[ht]
\caption{Position, distance, and J factor (under assumption of a Navarro-Frenk-White profile) of each dSph. The 4th column shows the mode of the posterior distribution of $\log_{10}J$, and the 5th column indicates its 68\% C.L. error. See the text for further details. The J factors correspond to the pair annihilation flux coming from a cone of solid angle $\Delta\Omega = 2.4\cdot10^{-4}$ sr. The final column indicates the reference for the kinematic dataset used. \label{dwarfs}} 
{\small \hfill{}
\begin{ruledtabular}
\begin{tabular}{ccccccc}
Name & l &  b & d & $\overline{\log_{10}({J})}$ & $\sigma$ & ref.\\  
     & deg.& deg.& kpc&\multicolumn{2}{c}{$\log_{10}[{\rm GeV}^2 {\rm cm}^{-5}]$} \\\hline
Bootes I & $358.08$ & $\phantom{-}69.62$ &60 &$17.7$ &$0.34$ &\cite{Koposov:2011zi} \\ 
Carina & $260.11$ & $-22.22$ & 101&$18.0$ &$0.13$ &\cite{Walker:2009zp} \\ 
Coma Berenices & $241.9$ &$\phantom{-}83.6$ &44& $19.0$ & $0.37$ &\cite{Simon:2007dq}\\ 
Draco & $86.37$ & $\phantom{-}34.72$ &80 &$18.8$ & $0.13$ &\cite{Walker:2009zp}\\ 
Fornax & $237.1$ & $-65.7$ & 138&$17.7$ & $0.23$ &\cite{Walker:2009zp}\\ 
Sculptor & $287.15$ & $-83.16$ & 80&$18.4$ &$0.13$ &\cite{Walker:2009zp} \\ 
Segue 1 & $220.48$ & $\phantom{-}50.42$ & 23&$19.6$ & $0.53$ &\cite{Simon:2010ek} \\ 
Sextans & $243.4$ & $\phantom{-}42.2$ &86 &$17.8$ & $0.23$ &\cite{Walker:2009zp}\\ 
Ursa Major II & $152.46$ & $\phantom{-}37.44$ & 32&$19.6$ &$0.40$ &\cite{Simon:2007dq}\\ 
Ursa Minor & $104.95$ & $\phantom{-}44.80$ & 66&$18.5$ &$0.18$ &\cite{Walker:2009zp} \\ 
\end{tabular}
\end{ruledtabular}
}
\hfill{}
\end{table}

In this Letter we add Segue 1 and Carina to the sample of 8 dSphs analyzed in \cite{Abdo:2010ex}, where further details on the selection criteria are provided. Carina has been added as two years of data now allow us to reasonably model the Galactic diffuse background at Galactic latitudes about $ -22^\circ$. Segue 1 has been added because there has been significant progress in estimating its DM distribution recently \cite{Simon:2010ek,Martinez:2010xn,Essig:2010em}; however, uncertainties in these estimates remain large.

This Letter uses the instrument response functions P6V3 \cite{Atwood:2009ez} for the \textquotedblleft diffuse\textquotedblright  class of events. We also use the diffuse emission model derived and recommended by the $Fermi$-LAT Collaboration \cite{modelPage}. It includes the Galactic diffuse emission component (\textit{gll$\_$iem$\_$v02.fit}), and a corresponding isotropic component (\textit{isotropic$\_$iem$\_$v02.txt}) that accounts for isotropic background light, unresolved sources and residual cosmic-ray contamination. Point sources from the 1FGL catalog \cite{Collaboration:2010ru} within 15$^\circ$ of each dSph (and a few additional faint sources detected in two years of data) are included in the model. A potential DM signal in each ROI is modeled as a point source where the gamma-ray yields are obtained from the DMFit package \cite{Jeltema:2008hf} based on DarkSUSY \cite{Gondolo:2004sc}, as implemented in the ScienceTools. For the J factors (defined in the Introduction), we use the updated values summarized in Table ~\ref{dwarfs}, which were estimated as described in the next section. 

\section{J factors from stellar velocity data}
\label{sec:stellar}
J factors are calculated using the line-of-sight velocities of the stars in the dSph and the Jeans equation via a Bayesian method as described in the literature ({\em e.g.}, \cite{Evans:2003sc,Strigari:2007at,Martinez:2009jh}). The mass of DM within the half-light radii of the dSphs is well-constrained, independent of the assumption of whether there is a core or a cusp in the central DM density distribution \cite{Walker:2009zp,Wolf:2009tu}. 

We assume that the inner DM density profile scales as $1/r$, a close match to the results seen in dark-matter-only simulations where the particles are initially cold (like WIMPs). Baryonic processes may alter the density profiles in dSphs \cite{Parry:2011iz}. However, present velocity data are unable to differentiate between cores and cusps in dSphs in a model-independent manner. If the dSphs have constant density  cores, then, in order to match the stellar velocity data constraint (essentially the dynamical mass within the half-light radius), the normalization of the density profile at the half-light radius would have to be increased (compared to the $1/r$ profile). For large constant density cores (comparable to or larger than the dSph half-light radius), this results in a larger J factor if the pair annihilation flux is integrated over a solid angle larger than that encompassing half the stellar luminosity. This is due to the fact that flux is dominated by annihilations in the outer parts for $1/r$ and shallower dark matter density profiles. For small cores, the J factor can be smaller but the change is proportionally smaller also.

The observed half-light radii of the dSphs is less than or close to $0.5^\circ$ (which is the radius corresponding to the solid angle of $\Delta\Omega = 2.4\cdot10^{-4}$ sr used to compute the J factor). Thus, if we were to adopt a cored dark matter profile, the J factors for most of the dSphs would either increase or not change much. We have not attempted to model the possible correlations in the J-factor estimates of the different dSphs that would arise due to common baryonic feedback processes in these systems. These processes could, for example, create large constant density cores in the dark matter halos of all the dSphs. With a deeper understanding of galaxy formation on these small scales, it may be possible to refine the present constraint.

The DM mass distribution as a function of the radius from the center of the dwarf is modeled as a Navarro-Frenk-White profile given by $\rho(r) = 0.08 \vmax^2 r_s/[G r (r+r_s)^2]$, where $\vmax$ is the maximum circular velocity possible for the dark matter halo. 
For this profile, the J factor in units of ${\rm GeV}^2 {\rm cm}^{-5}$ $\simeq 10^{17} (\vmax/10 \,{\rm km} \,{\rm s}^{-1})^4({\rm kpc}/r_s)(100 \,{\rm kpc}/d)^2$ up to a function of $(d/r_s)(\Delta \Omega/\pi)^{1/2}$ that is of the order of unity for parameters of interest.

The stellar velocities used in the calculations are taken from the references listed in Table \ref{dwarfs}. For the 6 classical dwarfs, we used the available velocity dispersion data in radial bins \cite{Walker:2009zp}, and, for the fainter dwarfs (discovered in SDSS), we used the individual stellar velocities \cite{Koposov:2011zi,Simon:2007dq}. We used a Gaussian distribution for the line-of-sight velocity measurements, adding intrinsic velocity dispersion and measurement error in quadrature (see Eq. 13 of \cite{Martinez:2009jh}) and imposed spherical symmetry. For the binned velocity dispersion data, we used an approximation starting with the same Gaussian distribution for velocities and then assuming that the intrinsic velocity dispersion dominates the average measurement error (see Eq. 14 of \cite{Martinez:2009jh}). From tests on a few dwarfs, we expect that this approximation could introduce a bias of about 50\% to the most probable individual J factors. Other approximate likelihoods we tested also resulted in similar biases compared to Gaussian distribution for velocities.
We assume a flat prior in $\ln(\vmax)$ and a prior for $r_s$ given $\vmax$ consistent with both Aquarius and Via Lactea II simulations \cite{Springel:2008cc,Kuhlen:2009kx}. For the dSphs with the highest-quality datasets ({\em i.e.} the ones with the most stars and smallest errors, including Draco and Ursa Minor), the results do not change significantly if the flat $\ln(\vmax)$ prior is changed to match the $\vmax$ distribution of subhalos in $\Lambda$ cold dark matter simulations \cite{Martinez:2009jh} or if we assume a flat $\ln(r_s)$ prior. However, for dSphs with sparse datasets, such as ultrafaint Ursa Major II (20 stars) and Segue 1 (66 stars), the results are prior dependent. For example, adopting the subhalo prior for $\vmax$ decreases the median $J$ by a factor of $\sim 2$ for Segue 1 and Ursa Major II. The ultrafaint dSphs are promising candidates, but these and other significant uncertainties remain in the estimates of their DM halo mass. Considerable progress in dealing with some of these uncertainties has been made for Segue 1 \cite{Simon:2010ek,Martinez:2010xn,Essig:2010em}, but we have opted to treat both Segue 1 and Ursa Major II in the same fashion as the other dSphs for the sake of uniformity in treating the priors. This is a limitation of the analysis at present, so we quote constraints with and without Segue 1 and Ursa Major II below. The final results for the J factors within $\Delta\Omega = 2.4 \cdot 10^{-4}$ sr are listed in Table \ref{dwarfs}. To be conservative, we assume no contribution to the flux from DM substructure in the dSphs. The posterior distribution as well as the likelihood function for $J$ are well-described by a log normal function, which is used in order to include the uncertainty on $J$ in the confidence interval calculation, as described in the next section. 

\section{Data Analysis}
The ScienceTools analysis package is used to perform a binned Poisson likelihood fit to both spatial and spectral information in the data, with 30 energy bins logarithmically spaced from 200 MeV to 100 GeV and $10^\circ$ square spatial maps with a bin size of $0.1^\circ$. The normalizations of the two diffuse components are left free in all ROIs, together with the normalizations of the point sources within $5^\circ$ of the dSph position. The first improvement to the analysis in \cite{Abdo:2010ex} consists of combining the DM signal across all the ROIs. Indeed, the J factor is different for each dSph, but the characteristics of the WIMP candidate ($m_W$, \sv, annihilation channels and their branching ratios) can be assumed to be universal. As a consequence, the $Fermi$-LAT Collaboration developed the \emph{Composite2} code in the ScienceTools, to allow tying any set of parameters across any set of ROIs. The second improvement is that uncertainties on the J factor are taken into account in the fit procedure by adding another term to the likelihood that represents the measurement uncertainties. With this addition, the joint likelihood considered in our analysis becomes
\begin{align}\label{eq:L}
L(D| \mathbf{p_W},&\{\mathbf{p}\}_i)=\prod_{i} L^{\rm LAT}_{i}(D|\mathbf{p_W}, \mathbf{p}_i)\nonumber\\
 & \times \frac{1}{\ln(10)\,J_i \sqrt{2\pi}\sigma_{i}} e^{-\left[\log_{10}(J_i)-\overline{\log_{10}({J}_i)}\right]^2/2\sigma_{i}^2}\; ,
\end{align}
\noindent where $L^{\rm LAT}_i$ denotes the binned Poisson likelihood that is commonly used in a standard single ROI analysis of the LAT data and takes full account of the point-spread function, including its energy dependence; $i$ indexes the ROIs; $D$ represents the binned gamma-ray data; $\mathbf{p_W}$ represents the set of ROI-independent DM parameters ($\sv$ and $m_W$); and $\{\mathbf{p}\}_i$ are the ROI-dependent model parameters. In this analysis, $\{\mathbf{p}\}_i$ includes the normalizations of the nearby point and diffuse sources and the J factor, $J_i$. $\overline{\log_{10}(J_i)}$ and $\sigma_i$ are the mean and standard deviations of the distribution of $\log_{10}{(J_i)}$, approximated to be Gaussian, and their values are given in Columns 5 and 6, respectively, of Table~\ref{dwarfs}.

The fit proceeds as follows. For given fixed values of $m_W$ and $\mathbf{b_f}$, we optimize $-\ln L$, with $L$ given in Eq.~\ref{eq:L}. Confidence intervals or upper limits, taking into account uncertainties in the nuisance parameters, are then computed using the  \textquotedblleft profile likelihood\textquotedblright technique, which is a standard method for treating nuisance parameters in likelihood analyses (see, {\em e.g.}, \cite{Rolke:2004mj}), and consists of calculating the profile likelihood $-\ln L_p$(\sv) for several fixed masses $m_W$, where, for each \sv , $-\ln L$ is minimized with respect to all other parameters. The intervals are then obtained by requiring $2\Delta\ln(L_p) = 2.71$ for a one-sided 95\% confidence level. The MINUIT subroutine MINOS \cite{minuit}  is used as the implementation of this technique. Note that uncertainties in the background fit (diffuse and nearby sources) are also treated in this way. To summarize, the free parameters of the fit are $\sv$, the J factors, and the Galactic diffuse and isotropic background normalizations as well as the normalizations of near-by point sources. The coverage of this profile joint likelihood method for calculating confidence intervals has been verified using toy Monte Carlo calculations for a Poisson process with known background and $Fermi$-LAT simulations of Galactic and isotropic diffuse gamma-ray emission. The parameter range for $\sv\,$ is restricted to have a lower bound of zero, to facilitate convergence of the MINOS fit, resulting in slight overcoverage for small signals, {\em i.e.}, conservative limits.  

\section{Results and Conclusions}
\begin{figure}
\includegraphics[width=\columnwidth]{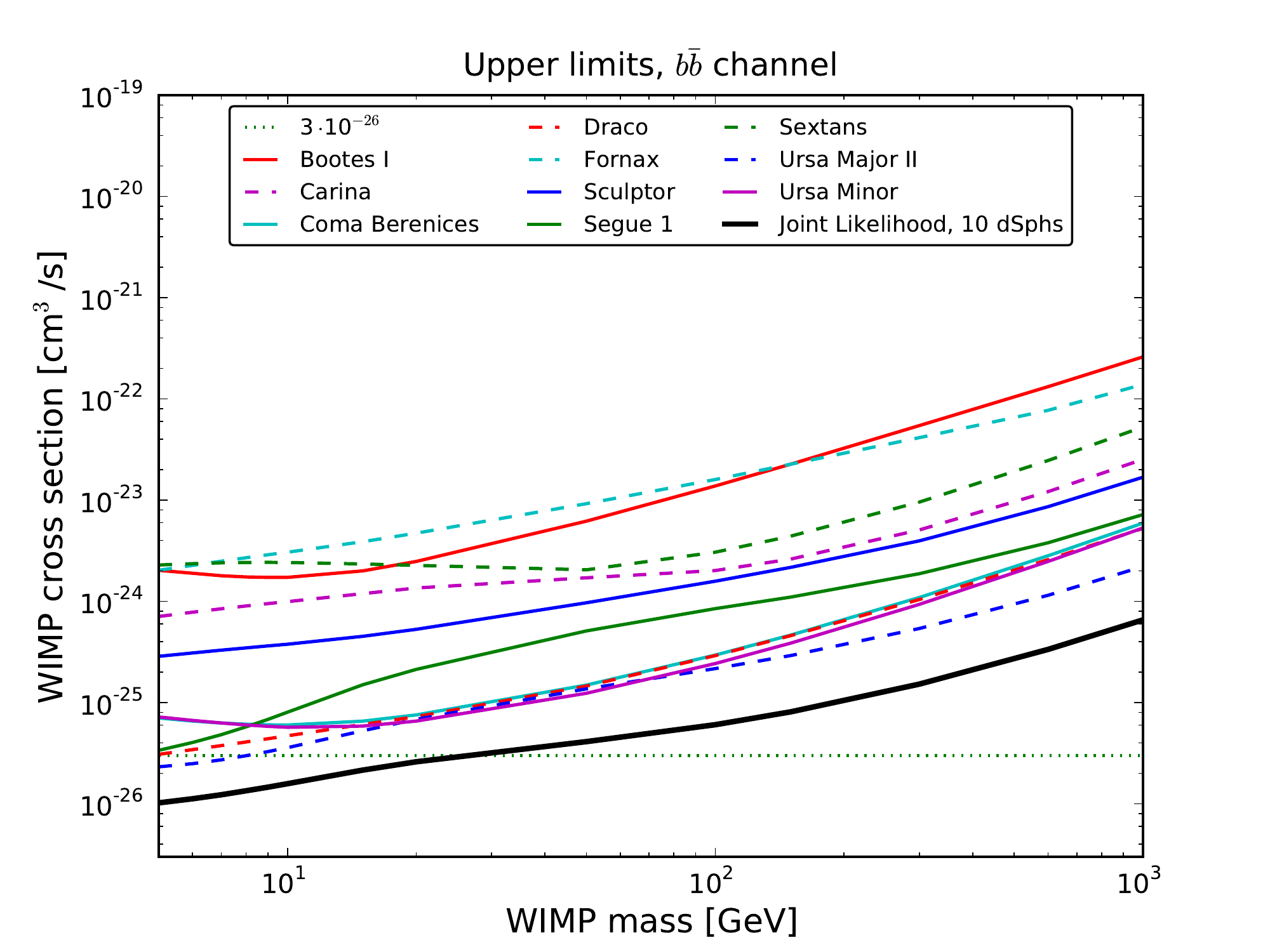}
\caption{Derived 95\% C.L. upper limits on a WIMP annihilation cross section for all selected dSphs and for the joint likelihood analysis for annihilation into the \textit{b\={b}} final state. The most generic cross section ($\sim 3 \cdot 10^{-26} \,{\rm cm}^3 {\rm s}^{-1}$ for a purely s-wave cross section) is plotted as a reference. Uncertainties in the J factor are included. \label{ULbbar}}
\end{figure}
\begin{figure}
\includegraphics[width=\columnwidth]{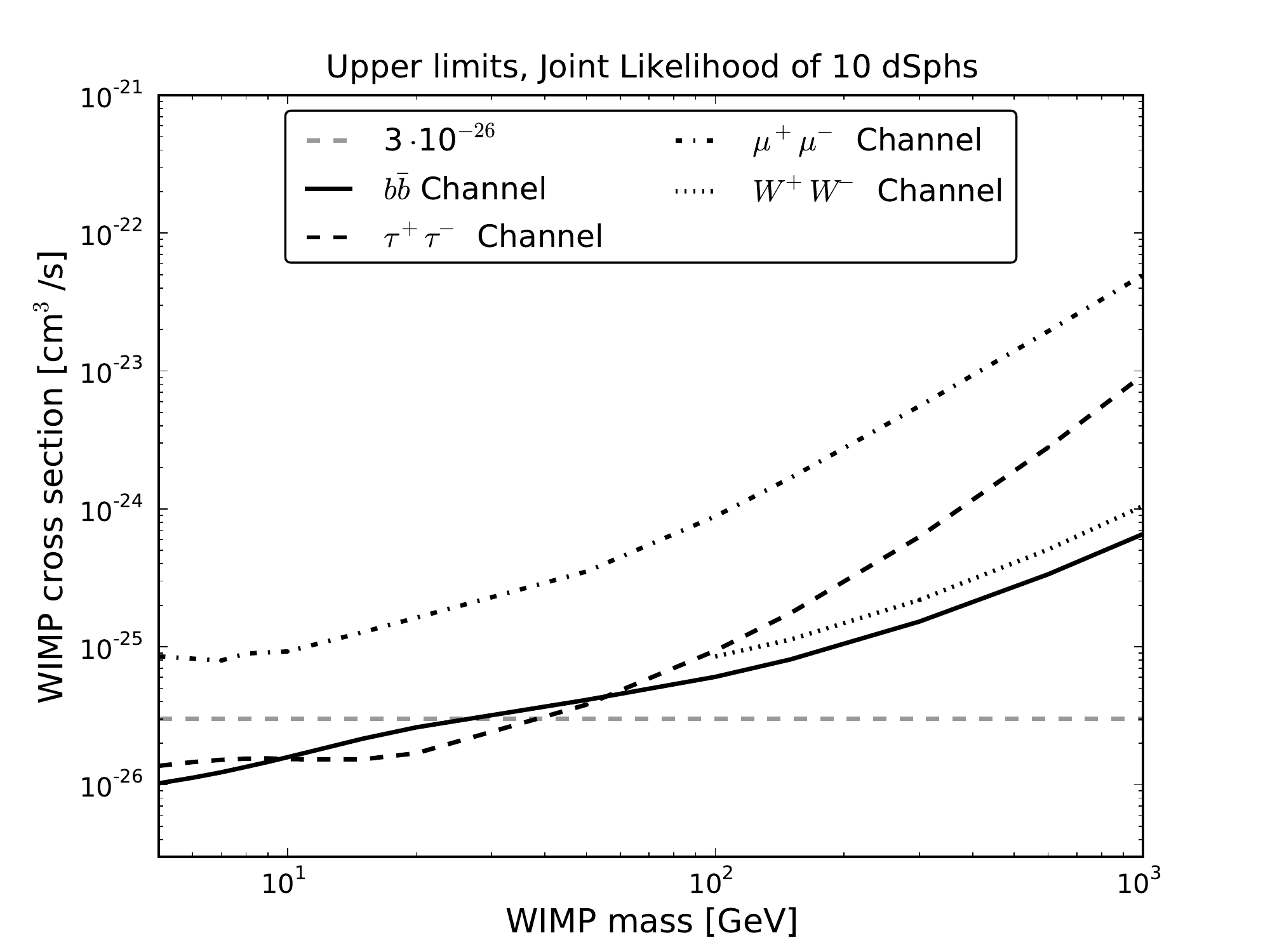}
\caption{Derived 95\% C.L. upper limits on a WIMP annihilation cross section for the \textit{b\={b}} channel, the $\tau^+ \tau^-$ channel, the $\mu^+ \mu^-$ channel, and the $W^+ W^-$ channel. The most generic cross section ($\sim 3 \cdot 10^{-26} \,{\rm cm}^3 {\rm s}^{-1}$ for a purely s-wave cross section) is plotted as a reference. Uncertainties in the J factor are included. \label{UL_J}}
\end{figure}
As no significant signal is found, we report upper limits. Individual and combined upper limits on the annihilation cross section for the \textit{b\={b}} final state are shown in Fig.~\ref{ULbbar}; see also \cite{supMaterial}. Including the J-factor uncertainties in the fit results in increased upper limits compared to using the nominal J factors. Averaged over the WIMP masses, the upper limits increase by a factor up to 12 for Segue 1, and down to 1.2 for Draco. Combining the dSphs yields a much milder overall increase of the upper limit compared to using nominal J factors, a factor of 1.3. 

The combined upper limit curve shown in Fig.~\ref{ULbbar} includes Segue 1 and Ursa Major II, two ultrafaint satellites with small kinematic data sets and relatively large uncertainties on their J factors. Conservatively, excluding these objects from the analysis results in an increase in the upper limit by a factor $\sim$1.5, which illustrates the robustness of the combined fit. 

We recalculated our combined limits using, for the classical dwarfs, the J factors presented in \cite{Charbonnier:2011ft}, which allow for shallower profiles than Navarro-Frenk-White assumed here. The final constraint agrees with the limit from our J factors to about $10\%$, demonstrating the insensitivity of the combined limits to the assumed dark matter density profile.

Finally, Fig.~\ref{UL_J} shows the combined limits for all studied channels. The WIMP masses range from 5 GeV to 1 TeV, except for the $W^+ W^-$ channel, where the lower bound is 100 GeV. For the first time, using gamma rays, we are able to rule out models with the most generic cross section ($\sim 3 \cdot 10^{-26} \,{\rm cm}^3 {\rm s}^{-1}$ for a purely s-wave cross section), without assuming additional astrophysical or particle physics boost factors. For large dark matter masses (around or above a TeV), the radiation of soft electro-weak bosons leads to additional gamma rays in the energy range of relevance for the present analysis (see, e.g., \cite{Bell:2008ey, Ciafaloni:2010ti}). This emission mechanism is not included in the Monte Carlo simulations for the photon yield we employ here. While massive gauge boson radiation is virtually irrelevant for masses below 100 GeV, our results for the heaviest masses can be instead viewed as marginally more conservative than with the inclusion of radiative electro-weak corrections.

In conclusion, we have presented a new analysis of the $Fermi$-LAT data that for the first time combines multiple (10) Milky Way satellite galaxies in a single joint likelihood fit and includes the effects of uncertainties in updated J factors, yielding a more robust upper limit curve in the ($m_W$,\sv) plane. This procedure allows us to rule out WIMP annihilation, with cross sections predicted by the most generic cosmological calculation up to a mass of $\sim 27$ GeV for the \textit{b\={b}} channel and up to a mass of $\sim 37$ GeV for the $\tau^+ \tau^-$ channel. Future improvements planned by the $Fermi$-LAT Collaboration (apart from an increased amount of data) will include an improved event selection with a larger effective area and photon energy range and the inclusion of more satellite galaxies.

\begin{acknowledgments}
The $Fermi$-LAT Collaboration acknowledges support from a number of agencies and institutes for both development and the operation of the LAT as well as scientific data analysis. These include NASA and DOE in the United States, CEA/Irfu and IN2P3/CNRS in France, ASI and INFN in Italy, MEXT, KEK, and JAXA in Japan, and the K.~A.~Wallenberg Foundation, the Swedish Research Council and the National Space Board in Sweden. Additional support from INAF in Italy and CNES in France for science analysis during the operations phase is also gratefully acknowledged. External collaborators M. Kaplinghat and G. D. Martinez acknowledge support from NASA grant NNX09AD09G.
\end{acknowledgments}

\emph{Note added in proof.-} During the final preparation for submission of this Letter, we became aware of the work by Geringer-Sameth and Koushiappas when it was posted to the arXiv \cite{Geringer}, reaching similar conclusions as ours, albeit with a different analysis.

\bibliography{bibl_compLikePaper}

\providecommand{\noopsort}[1]{}\providecommand{\singleletter}[1]{#1}%
\begin{thebibliography}{38}%
\makeatletter
\providecommand \@ifxundefined [1]{%
 \@ifx{#1\undefined}
}%
\providecommand \@ifnum [1]{%
 \ifnum #1\expandafter \@firstoftwo
 \else \expandafter \@secondoftwo
 \fi
}%
\providecommand \@ifx [1]{%
 \ifx #1\expandafter \@firstoftwo
 \else \expandafter \@secondoftwo
 \fi
}%
\providecommand \natexlab [1]{#1}%
\providecommand \enquote  [1]{``#1''}%
\providecommand \bibnamefont  [1]{#1}%
\providecommand \bibfnamefont [1]{#1}%
\providecommand \citenamefont [1]{#1}%
\providecommand \href@noop [0]{\@secondoftwo}%
\providecommand \href [0]{\begingroup \@sanitize@url \@href}%
\providecommand \@href[1]{\@@startlink{#1}\@@href}%
\providecommand \@@href[1]{\endgroup#1\@@endlink}%
\providecommand \@sanitize@url [0]{\catcode `\\12\catcode `\$12\catcode
  `\&12\catcode `\#12\catcode `\^12\catcode `\_12\catcode `\%12\relax}%
\providecommand \@@startlink[1]{}%
\providecommand \@@endlink[0]{}%
\providecommand \url  [0]{\begingroup\@sanitize@url \@url }%
\providecommand \@url [1]{\endgroup\@href {#1}{\urlprefix }}%
\providecommand \urlprefix  [0]{URL }%
\providecommand \Eprint [0]{\href }%
\@ifxundefined \urlstyle {%
  \providecommand \doi  [0]{\begingroup \@sanitize@url \@doi}%
  \providecommand \@doi [1]{\endgroup \@@startlink {\doibase
  #1}doi:\discretionary {}{}{}#1\@@endlink }%
}{%
  \providecommand \doi  [0]{doi:\discretionary{}{}{}\begingroup
  \urlstyle{rm}\Url }%
}%
\providecommand \doibase [0]{http://dx.doi.org/}%
\providecommand \Doi [0]{\begingroup \@sanitize@url \@Doi }%
\providecommand \@Doi  [1]{\endgroup\@@startlink{\doibase#1}\@@Doi}%
\providecommand \@@Doi [1]{#1\@@endlink}%
\providecommand \selectlanguage [0]{\@gobble}%
\providecommand \bibinfo  [0]{\@secondoftwo}%
\providecommand \bibfield  [0]{\@secondoftwo}%
\providecommand \translation [1]{[#1]}%
\providecommand \BibitemOpen [0]{}%
\providecommand \bibitemStop [0]{}%
\providecommand \bibitemNoStop [0]{.\EOS\space}%
\providecommand \EOS [0]{\spacefactor3000\relax}%
\providecommand \BibitemShut  [1]{\csname bibitem#1\endcsname}%
\bibitem [{\citenamefont {Komatsu}\ \emph {et~al.}(2011)\citenamefont {Komatsu}
  \emph {et~al.}}]{Komatsu:2010fb}%
  \BibitemOpen
  \bibfield  {author} {\bibinfo {author} {\bibfnamefont {E.}~\bibnamefont
  {Komatsu}} \emph {et~al.} (\bibinfo {collaboration} {WMAP}),\ }\Doi
  {10.1088/0067-0049/192/2/18} {\bibfield  {journal} {\bibinfo  {journal}
  {Astrophys. J. Suppl.},\ }\textbf {\bibinfo {volume} {192}},\ \bibinfo
  {pages} {18} (\bibinfo {year} {2011})}\BibitemShut {NoStop}%
\bibitem [{\citenamefont {Bergstrom}\ \emph {et~al.}(1998)\citenamefont
  {Bergstrom}, \citenamefont {Ullio},\ and\ \citenamefont
  {Buckley}}]{Bergstrom:1997fj}%
  \BibitemOpen
  \bibfield  {author} {\bibinfo {author} {\bibfnamefont {L.}~\bibnamefont
  {Bergstrom}}, \bibinfo {author} {\bibfnamefont {P.}~\bibnamefont {Ullio}}, \
  and\ \bibinfo {author} {\bibfnamefont {J.~H.}\ \bibnamefont {Buckley}},\
  }\Doi {10.1016/S0927-6505(98)00015-2} {\bibfield  {journal} {\bibinfo
  {journal} {Astropart.Phys.},\ }\textbf {\bibinfo {volume} {9}},\ \bibinfo
  {pages} {137} (\bibinfo {year} {1998})}\BibitemShut {NoStop}%
\bibitem [{\citenamefont {Bergstrom}(2000)}]{Bergstrom:2000pn}%
  \BibitemOpen
  \bibfield  {author} {\bibinfo {author} {\bibfnamefont {L.}~\bibnamefont
  {Bergstrom}},\ }\Doi {10.1088/0034-4885/63/5/2r3} {\bibfield  {journal}
  {\bibinfo  {journal} {Rept.Prog.Phys.},\ }\textbf {\bibinfo {volume} {63}},\
  \bibinfo {pages} {793} (\bibinfo {year} {2000})}\BibitemShut {NoStop}%
\bibitem [{\citenamefont {Mateo}(1998)}]{Mateo:1998wg}%
  \BibitemOpen
  \bibfield  {author} {\bibinfo {author} {\bibfnamefont {M.}~\bibnamefont
  {Mateo}},\ }\Doi {10.1146/annurev.astro.36.1.435} {\bibfield  {journal}
  {\bibinfo  {journal} {Ann. Rev. Astron. Astrophys.},\ }\textbf {\bibinfo
  {volume} {36}},\ \bibinfo {pages} {435} (\bibinfo {year} {1998})}\BibitemShut
  {NoStop}%
\bibitem [{\citenamefont {Grcevich}\ and\ \citenamefont
  {Putman}(2009)}]{Grcevich:2009gt}%
  \BibitemOpen
  \bibfield  {author} {\bibinfo {author} {\bibfnamefont {J.}~\bibnamefont
  {Grcevich}}\ and\ \bibinfo {author} {\bibfnamefont {M.~E.}\ \bibnamefont
  {Putman}},\ }\Doi {10.1088/0004-637X/696/1/385} {\bibfield  {journal}
  {\bibinfo  {journal} {Astrophys. J.},\ }\textbf {\bibinfo {volume} {696}},\
  \bibinfo {pages} {385} (\bibinfo {year} {2009})}\BibitemShut {NoStop}%
\bibitem [{\citenamefont {Abdo}\ \emph
  {et~al.}(2010){\natexlab{a}}\citenamefont {Abdo} \emph
  {et~al.}}]{Abdo:2010ex}%
  \BibitemOpen
  \bibfield  {author} {\bibinfo {author} {\bibfnamefont {A.~A.}\ \bibnamefont
  {Abdo}} \emph {et~al.} (\bibinfo {collaboration} {LAT}),\ }\Doi
  {10.1088/0004-637X/712/1/147} {\bibfield  {journal} {\bibinfo  {journal}
  {Astrophys. J.},\ }\textbf {\bibinfo {volume} {712}},\ \bibinfo {pages} {147}
  (\bibinfo {year} {2010}{\natexlab{a}})}\BibitemShut {NoStop}%
\bibitem [{\citenamefont {Scott}\ \emph {et~al.}(2010)\citenamefont {Scott}
  \emph {et~al.}}]{Scott:2009jn}%
  \BibitemOpen
  \bibfield  {author} {\bibinfo {author} {\bibfnamefont {P.}~\bibnamefont
  {Scott}} \emph {et~al.},\ }\Doi {10.1088/1475-7516/2010/01/031} {\bibfield
  {journal} {\bibinfo  {journal} {JCAP},\ }\textbf {\bibinfo {volume} {1001}},\
  \bibinfo {pages} {031} (\bibinfo {year} {2010})}\BibitemShut {NoStop}%
\bibitem [{\citenamefont {Aharonian}\ \emph {et~al.}(2008)\citenamefont
  {Aharonian} \emph {et~al.}}]{Aharonian:2007km}%
  \BibitemOpen
  \bibfield  {author} {\bibinfo {author} {\bibfnamefont {F.}~\bibnamefont
  {Aharonian}} \emph {et~al.} (\bibinfo {collaboration} {HESS}),\ }\Doi
  {10.1016/j.astropartphys.2007.11.007} {\bibfield  {journal} {\bibinfo
  {journal} {Astropart. Phys.},\ }\textbf {\bibinfo {volume} {29}},\ \bibinfo
  {pages} {55} (\bibinfo {year} {2008})}\BibitemShut {NoStop}%
\bibitem [{\citenamefont {Albert}\ \emph {et~al.}(2008)\citenamefont {Albert}
  \emph {et~al.}}]{Albert:2007xg}%
  \BibitemOpen
  \bibfield  {author} {\bibinfo {author} {\bibfnamefont {J.}~\bibnamefont
  {Albert}} \emph {et~al.} (\bibinfo {collaboration} {MAGIC}),\ }\Doi
  {10.1086/529135} {\bibfield  {journal} {\bibinfo  {journal} {Astrophys. J.},\
  }\textbf {\bibinfo {volume} {679}},\ \bibinfo {pages} {428} (\bibinfo {year}
  {2008})}\BibitemShut {NoStop}%
\bibitem [{\citenamefont {Acciari}\ \emph {et~al.}(2010)\citenamefont {Acciari}
  \emph {et~al.}}]{:2010pja}%
  \BibitemOpen
  \bibfield  {author} {\bibinfo {author} {\bibfnamefont {V.~A.}\ \bibnamefont
  {Acciari}} \emph {et~al.} (\bibinfo {collaboration} {VERITAS}),\ }\Doi
  {10.1088/0004-637X/720/2/1174} {\bibfield  {journal} {\bibinfo  {journal}
  {Astrophys. J.},\ }\textbf {\bibinfo {volume} {720}},\ \bibinfo {pages}
  {1174} (\bibinfo {year} {2010})}\BibitemShut {NoStop}%
\bibitem [{\citenamefont {Aleksic}\ \emph {et~al.}(2011)\citenamefont {Aleksic}
  \emph {et~al.}}]{Aleksic:2011jx}%
  \BibitemOpen
  \bibfield  {author} {\bibinfo {author} {\bibfnamefont {J.}~\bibnamefont
  {Aleksic}} \emph {et~al.} (\bibinfo {collaboration} {MAGIC}),\ }\Doi
  {10.1088/1475-7516/2011/06/035} {\bibfield  {journal} {\bibinfo  {journal}
  {JCAP},\ }\textbf {\bibinfo {volume} {1106}},\ \bibinfo {pages} {035}
  (\bibinfo {year} {2011})}\BibitemShut {NoStop}%
\bibitem [{\citenamefont {Atwood}\ \emph {et~al.}(2009)\citenamefont {Atwood}
  \emph {et~al.}}]{Atwood:2009ez}%
  \BibitemOpen
  \bibfield  {author} {\bibinfo {author} {\bibfnamefont {W.~B.}\ \bibnamefont
  {Atwood}} \emph {et~al.} (\bibinfo {collaboration} {LAT}),\ }\Doi
  {10.1088/0004-637X/697/2/1071} {\bibfield  {journal} {\bibinfo  {journal}
  {Astrophys. J.},\ }\textbf {\bibinfo {volume} {697}},\ \bibinfo {pages}
  {1071} (\bibinfo {year} {2009})}\BibitemShut {NoStop}%
\bibitem [{Per()}]{Perfpage}%
  \BibitemOpen
  \href@noop {} {}\bibinfo {howpublished}
  {http://www-glast.slac.stanford.edu/software/IS/ \\
  glast\_lat\_performance.htm}\BibitemShut {NoStop}%
\bibitem [{STp()}]{STpage}%
  \BibitemOpen
  \href@noop {} {}\bibinfo {howpublished}
  {http://fermi.gsfc.nasa.gov/ssc/data/analysis/software/}\BibitemShut
  {NoStop}%
\bibitem [{\citenamefont {Koposov}\ \emph {et~al.}(2011)\citenamefont {Koposov}
  \emph {et~al.}}]{Koposov:2011zi}%
  \BibitemOpen
  \bibfield  {author} {\bibinfo {author} {\bibfnamefont {S.~E.}\ \bibnamefont
  {Koposov}} \emph {et~al.},\ }\Doi {10.1088/0004-637X/736/2/146} {\bibfield
  {journal} {\bibinfo  {journal} {Astrophys.J.},\ }\textbf {\bibinfo {volume}
  {736}},\ \bibinfo {pages} {146} (\bibinfo {year} {2011})}\BibitemShut
  {NoStop}%
\bibitem [{\citenamefont {Walker}\ \emph {et~al.}(2009)\citenamefont {Walker}
  \emph {et~al.}}]{Walker:2009zp}%
  \BibitemOpen
  \bibfield  {author} {\bibinfo {author} {\bibfnamefont {M.~G.}\ \bibnamefont
  {Walker}} \emph {et~al.},\ }\Doi {10.1088/0004-637X/704/2/1274,
  10.1088/0004-637X/710/1/886} {\bibfield  {journal} {\bibinfo  {journal}
  {Astrophys.J.},\ }\textbf {\bibinfo {volume} {704}},\ \bibinfo {pages} {1274}
  (\bibinfo {year} {2009})}\BibitemShut {NoStop}%
\bibitem [{\citenamefont {Simon}\ and\ \citenamefont
  {Geha}(2007)}]{Simon:2007dq}%
  \BibitemOpen
  \bibfield  {author} {\bibinfo {author} {\bibfnamefont {J.~D.}\ \bibnamefont
  {Simon}}\ and\ \bibinfo {author} {\bibfnamefont {M.}~\bibnamefont {Geha}},\
  }\Doi {10.1086/521816} {\bibfield  {journal} {\bibinfo  {journal}
  {Astrophys.J.},\ }\textbf {\bibinfo {volume} {670}},\ \bibinfo {pages} {313}
  (\bibinfo {year} {2007})}\BibitemShut {NoStop}%
\bibitem [{\citenamefont {Simon}\ \emph {et~al.}(2011)\citenamefont {Simon}
  \emph {et~al.}}]{Simon:2010ek}%
  \BibitemOpen
  \bibfield  {author} {\bibinfo {author} {\bibfnamefont {J.~D.}\ \bibnamefont
  {Simon}} \emph {et~al.},\ }\Doi {10.1088/0004-637X/733/1/46} {\bibfield
  {journal} {\bibinfo  {journal} {Astrophys.J.},\ }\textbf {\bibinfo {volume}
  {733}},\ \bibinfo {pages} {46} (\bibinfo {year} {2011})}\BibitemShut
  {NoStop}%
\bibitem [{\citenamefont {Martinez}\ \emph {et~al.}(2011)\citenamefont
  {Martinez} \emph {et~al.}}]{Martinez:2010xn}%
  \BibitemOpen
  \bibfield  {author} {\bibinfo {author} {\bibfnamefont {G.~D.}\ \bibnamefont
  {Martinez}} \emph {et~al.},\ }\Doi {doi:10.1088/0004-637X/738/1/55}
  {\bibfield  {journal} {\bibinfo  {journal} {Astrophys.J.},\ }\textbf
  {\bibinfo {volume} {738}},\ \bibinfo {pages} {55} (\bibinfo {year}
  {2011})}\BibitemShut {NoStop}%
\bibitem [{\citenamefont {Essig}\ \emph {et~al.}(2010)\citenamefont {Essig}
  \emph {et~al.}}]{Essig:2010em}%
  \BibitemOpen
  \bibfield  {author} {\bibinfo {author} {\bibfnamefont {R.}~\bibnamefont
  {Essig}} \emph {et~al.},\ }\Doi {10.1103/PhysRevD.82.123503} {\bibfield
  {journal} {\bibinfo  {journal} {Phys.Rev.},\ }\textbf {\bibinfo {volume}
  {D82}},\ \bibinfo {pages} {123503} (\bibinfo {year} {2010})}\BibitemShut
  {NoStop}%
\bibitem [{mod()}]{modelPage}%
  \BibitemOpen
  \href@noop {} {}\bibinfo {howpublished}
  {http://fermi.gsfc.nasa.gov/ssc/data/access/lat/ \\
  BackgroundModels.html}\BibitemShut {NoStop}%
\bibitem [{\citenamefont {Abdo}\ \emph
  {et~al.}(2010){\natexlab{b}}\citenamefont {Abdo} \emph
  {et~al.}}]{Collaboration:2010ru}%
  \BibitemOpen
  \bibfield  {author} {\bibinfo {author} {\bibfnamefont {A.~A.}\ \bibnamefont
  {Abdo}} \emph {et~al.} (\bibinfo {collaboration} {LAT}),\ }\Doi
  {10.1088/0067-0049/188/2/405} {\bibfield  {journal} {\bibinfo  {journal}
  {Astrophys. J. Suppl.},\ }\textbf {\bibinfo {volume} {188}},\ \bibinfo
  {pages} {405} (\bibinfo {year} {2010}{\natexlab{b}})}\BibitemShut {NoStop}%
\bibitem [{\citenamefont {Jeltema}\ and\ \citenamefont
  {Profumo}(2008)}]{Jeltema:2008hf}%
  \BibitemOpen
  \bibfield  {author} {\bibinfo {author} {\bibfnamefont {T.~E.}\ \bibnamefont
  {Jeltema}}\ and\ \bibinfo {author} {\bibfnamefont {S.}~\bibnamefont
  {Profumo}},\ }\Doi {10.1088/1475-7516/2008/11/003} {\bibfield  {journal}
  {\bibinfo  {journal} {JCAP},\ }\textbf {\bibinfo {volume} {0811}},\ \bibinfo
  {pages} {003} (\bibinfo {year} {2008})}\BibitemShut {NoStop}%
\bibitem [{\citenamefont {Gondolo}\ \emph {et~al.}(2004)\citenamefont {Gondolo}
  \emph {et~al.}}]{Gondolo:2004sc}%
  \BibitemOpen
  \bibfield  {author} {\bibinfo {author} {\bibfnamefont {P.}~\bibnamefont
  {Gondolo}} \emph {et~al.},\ }\Doi {10.1088/1475-7516/2004/07/008} {\bibfield
  {journal} {\bibinfo  {journal} {JCAP},\ }\textbf {\bibinfo {volume} {0407}},\
  \bibinfo {pages} {008} (\bibinfo {year} {2004})}\BibitemShut {NoStop}%
\bibitem [{\citenamefont {Evans}\ \emph {et~al.}(2004)\citenamefont {Evans},
  \citenamefont {Ferrer},\ and\ \citenamefont {Sarkar}}]{Evans:2003sc}%
  \BibitemOpen
  \bibfield  {author} {\bibinfo {author} {\bibfnamefont {N.}~\bibnamefont
  {Evans}}, \bibinfo {author} {\bibfnamefont {F.}~\bibnamefont {Ferrer}}, \
  and\ \bibinfo {author} {\bibfnamefont {S.}~\bibnamefont {Sarkar}},\ }\Doi
  {10.1103/PhysRevD.69.123501} {\bibfield  {journal} {\bibinfo  {journal}
  {Phys.Rev.},\ }\textbf {\bibinfo {volume} {D69}},\ \bibinfo {pages} {123501}
  (\bibinfo {year} {2004})}\BibitemShut {NoStop}%
\bibitem [{\citenamefont {Strigari}\ \emph {et~al.}(2008)\citenamefont
  {Strigari} \emph {et~al.}}]{Strigari:2007at}%
  \BibitemOpen
  \bibfield  {author} {\bibinfo {author} {\bibfnamefont {L.~E.}\ \bibnamefont
  {Strigari}} \emph {et~al.},\ }\href@noop {} {\bibfield  {journal} {\bibinfo
  {journal} {\apj},\ }\textbf {\bibinfo {volume} {678}},\ \bibinfo {pages}
  {614} (\bibinfo {year} {2008})}\BibitemShut {NoStop}%
\bibitem [{\citenamefont {Martinez}\ \emph {et~al.}(2009)\citenamefont
  {Martinez} \emph {et~al.}}]{Martinez:2009jh}%
  \BibitemOpen
  \bibfield  {author} {\bibinfo {author} {\bibfnamefont {G.~D.}\ \bibnamefont
  {Martinez}} \emph {et~al.},\ }\Doi {10.1088/1475-7516/2009/06/014} {\bibfield
   {journal} {\bibinfo  {journal} {JCAP},\ }\textbf {\bibinfo {volume}
  {0906}},\ \bibinfo {pages} {014} (\bibinfo {year} {2009})}\BibitemShut
  {NoStop}%
\bibitem [{\citenamefont {Wolf}\ \emph {et~al.}(2010)\citenamefont {Wolf} \emph
  {et~al.}}]{Wolf:2009tu}%
  \BibitemOpen
  \bibfield  {author} {\bibinfo {author} {\bibfnamefont {J.}~\bibnamefont
  {Wolf}} \emph {et~al.},\ }\href@noop {} {\bibfield  {journal} {\bibinfo
  {journal} {Mon.Not.Roy.Astron.Soc.},\ }\textbf {\bibinfo {volume} {406}},\
  \bibinfo {pages} {1220} (\bibinfo {year} {2010})}\BibitemShut {NoStop}%
\bibitem [{\citenamefont {Parry}\ \emph {et~al.}(2011)\citenamefont {Parry}
  \emph {et~al.}}]{Parry:2011iz}%
  \BibitemOpen
  \bibfield  {author} {\bibinfo {author} {\bibfnamefont {O.~H.}\ \bibnamefont
  {Parry}} \emph {et~al.},\ }\href@noop {} { (\bibinfo {year} {2011})},\
  \Eprint {http://arxiv.org/abs/1105.3474} {arXiv:1105.3474} \BibitemShut
  {NoStop}%
\bibitem [{\citenamefont {Springel}\ \emph {et~al.}(2008)\citenamefont
  {Springel} \emph {et~al.}}]{Springel:2008cc}%
  \BibitemOpen
  \bibfield  {author} {\bibinfo {author} {\bibfnamefont {V.}~\bibnamefont
  {Springel}} \emph {et~al.},\ }\Doi {10.1111/j.1365-2966.2008.14066.x}
  {\bibfield  {journal} {\bibinfo  {journal} {Mon.Not.Roy.Astron.Soc.},\
  }\textbf {\bibinfo {volume} {391}},\ \bibinfo {pages} {1685} (\bibinfo {year}
  {2008})}\BibitemShut {NoStop}%
\bibitem [{\citenamefont {Kuhlen}\ \emph {et~al.}(2009)\citenamefont {Kuhlen},
  \citenamefont {Madau},\ and\ \citenamefont {Silk}}]{Kuhlen:2009kx}%
  \BibitemOpen
  \bibfield  {author} {\bibinfo {author} {\bibfnamefont {M.}~\bibnamefont
  {Kuhlen}}, \bibinfo {author} {\bibfnamefont {P.}~\bibnamefont {Madau}}, \
  and\ \bibinfo {author} {\bibfnamefont {J.}~\bibnamefont {Silk}},\ }\Doi
  {10.1126/science.1174881, 10.1126/science.1174881} {\bibfield  {journal}
  {\bibinfo  {journal} {Science},\ }\textbf {\bibinfo {volume} {325}},\
  \bibinfo {pages} {970} (\bibinfo {year} {2009})}\BibitemShut {NoStop}%
\bibitem [{\citenamefont {Rolke}\ \emph {et~al.}(2005)\citenamefont {Rolke},
  \citenamefont {Lopez},\ and\ \citenamefont {Conrad}}]{Rolke:2004mj}%
  \BibitemOpen
  \bibfield  {author} {\bibinfo {author} {\bibfnamefont {W.~A.}\ \bibnamefont
  {Rolke}}, \bibinfo {author} {\bibfnamefont {A.~M.}\ \bibnamefont {Lopez}}, \
  and\ \bibinfo {author} {\bibfnamefont {J.}~\bibnamefont {Conrad}},\ }\Doi
  {10.1016/j.nima.2005.05.068} {\bibfield  {journal} {\bibinfo  {journal}
  {Nucl.Instrum.Meth.},\ }\textbf {\bibinfo {volume} {A551}},\ \bibinfo {pages}
  {493} (\bibinfo {year} {2005})}\BibitemShut {NoStop}%
\bibitem [{\citenamefont {James}(1998)}]{minuit}%
  \BibitemOpen
  \bibfield  {author} {\bibinfo {author} {\bibfnamefont {F.}~\bibnamefont
  {James}},\ }\href@noop {} {\emph {\bibinfo {title} {MINUIT Reference
  Manual}}},\ \bibinfo {organization} {Version 94.1, CERN Program Library
  Writeup D506} (\bibinfo {year} {1998})\BibitemShut {NoStop}%
\bibitem [{sup()}]{supMaterial}%
  \BibitemOpen
  \href@noop {} {}\bibinfo {note} {Supplemental Material at
  http://link.aps.org/ supplemental/10.1103/PhysRevLett.107.241302 for
  numerical tables of the ULs}\BibitemShut {NoStop}%
\bibitem [{\citenamefont {Charbonnier}\ \emph {et~al.}(2011)\citenamefont
  {Charbonnier} \emph {et~al.}}]{Charbonnier:2011ft}%
  \BibitemOpen
  \bibfield  {author} {\bibinfo {author} {\bibfnamefont {A.}~\bibnamefont
  {Charbonnier}} \emph {et~al.},\ }\href@noop {} { (\bibinfo {year} {2011})},\
  \bibinfo {note} {accepted for publication in Mon.Not.Roy.Astron.Soc.},\
  \Eprint {http://arxiv.org/abs/1104.0412} {arXiv:1104.0412} \BibitemShut
  {NoStop}%
\bibitem [{\citenamefont {Bell}\ \emph {et~al.}(2008)\citenamefont {Bell} \emph
  {et~al.}}]{Bell:2008ey}%
  \BibitemOpen
  \bibfield  {author} {\bibinfo {author} {\bibfnamefont {N.~F.}\ \bibnamefont
  {Bell}} \emph {et~al.},\ }\Doi {10.1103/PhysRevD.78.083540} {\bibfield
  {journal} {\bibinfo  {journal} {Phys.Rev.},\ }\textbf {\bibinfo {volume}
  {D78}},\ \bibinfo {pages} {083540} (\bibinfo {year} {2008})}\BibitemShut
  {NoStop}%
\bibitem [{\citenamefont {Ciafaloni}\ \emph {et~al.}(2011)\citenamefont
  {Ciafaloni} \emph {et~al.}}]{Ciafaloni:2010ti}%
  \BibitemOpen
  \bibfield  {author} {\bibinfo {author} {\bibfnamefont {P.}~\bibnamefont
  {Ciafaloni}} \emph {et~al.},\ }\Doi {10.1088/1475-7516/2011/03/019}
  {\bibfield  {journal} {\bibinfo  {journal} {JCAP},\ }\textbf {\bibinfo
  {volume} {1103}},\ \bibinfo {pages} {019} (\bibinfo {year}
  {2011})}\BibitemShut {NoStop}%
\bibitem [{\citenamefont {Geringer-Sameth}\ and\ \citenamefont
  {Koushiappas}(2011)}]{Geringer}%
  \BibitemOpen
  \bibfield  {author} {\bibinfo {author} {\bibfnamefont {A.}~\bibnamefont
  {Geringer-Sameth}}\ and\ \bibinfo {author} {\bibfnamefont {S.~M.}\
  \bibnamefont {Koushiappas}},\ }\href@noop {} {\bibfield  {journal} {\bibinfo
  {journal} {following Letter, Phys. Rev. Lett.},\ }\textbf {\bibinfo {volume}
  {107}},\ \bibinfo {pages} {241303} (\bibinfo {year} {2011})}\BibitemShut
  {NoStop}%
\end{thebibliography}%

\end{document}